%% file: conference_101719.tex
\documentclass[conference]{IEEEtran}
\usepackage{cite}
\usepackage{amsmath, amssymb, amsfonts}
\usepackage{graphicx}
\usepackage{textcomp}
\usepackage{booktabs}
\usepackage{threeparttable}
\usepackage{tabularx}
\usepackage{xcolor}
\usepackage{multirow}
\usepackage{soul}
\usepackage{todonotes}
\usepackage{times}
\usepackage{latexsym}
\usepackage[T1]{fontenc}
\usepackage[utf8]{inputenc}
\usepackage{microtype}
\usepackage{enumitem}
\usepackage{inconsolata}
\usepackage{pifont}
\usepackage{float}
\usepackage{afterpage}
\usepackage{placeins}
\usepackage{adjustbox}
\usepackage{newunicodechar}
\usepackage{graphicx}      
\usepackage{subcaption} 
\usepackage{framed}
\newunicodechar{≥}{$\geq$}
\usepackage[framemethod=TikZ]{mdframed}
\usepackage{multicol}
\usepackage[most]{tcolorbox}
\usepackage[linesnumbered,ruled,vlined]{algorithm2e} 
\usepackage{hyperref}
\IEEEoverridecommandlockouts
\def\BibTeX{{\rm B\kern-.05em{\sc i\kern-.025em b}\kern-.08em
    T\kern-.1667em\lower.7ex\hbox{E}\kern-.125emX}}
\begin{document}

\title{TigerGPT: A New AI Chatbot for Adaptive Campus Climate Surveys\\}

\author{\IEEEauthorblockN{Jinwen Tang, Songxi Chen, Yi Shang}
\IEEEauthorblockA{\textit{Electrical Engineering and Computer Science Department} \\
\textit{University of Missouri}\\
Columbia, MO, USA \\
\{jt4cc, sc8rg, shangy\}@umsystem.edu}
}

\maketitle

\begin{abstract} 
Campus climate surveys play a pivotal role in capturing how students, faculty, and staff experience university life, yet traditional methods frequently suffer from low participation and minimal follow-up. We present TigerGPT, a new AI chatbot that generates adaptive, context-aware dialogues enriched with visual elements. Through real-time follow-up prompts, empathetic messaging, and flexible topic selection, TigerGPT elicits more in-depth feedback compared to traditional static survey forms. 
Based on established principles of conversational design, the chatbot employs empathetic
cues, bolded questions, and user-driven topic selection.
It 
retains some role-based efficiency (e.g., collecting user role through quick clicks) but goes beyond static scripts by employing GenAI adaptiveness.
In a pilot study with undergraduate students, we collected both quantitative metrics (e.g., satisfaction ratings) and qualitative insights (e.g., written comments). Most participants described TigerGPT as engaging and user-friendly; about half preferred it over conventional surveys, attributing this preference to its personalized conversation flow and supportive tone. The findings indicate that an AI survey chatbot is promising in gaining deeper insight into campus climate.
\end{abstract}


\section{Introduction}
Campus climate surveys play a crucial role in understanding how students, faculty, and staff experience everyday life at universities. By gathering feedback on topics ranging from academic support to discrimination and resource accessibility, these surveys help institutions identify structural barriers and develop strategies that foster equity and belonging \cite{vogel2008assessment, PASSHE2022, bordieri2024exploring}. Regularly collecting such feedback also enables ongoing monitoring of institutional progress, ensuring that efforts to improve campus environments remain both data-driven and responsive to community needs.

However, traditional surveys often fail to capture a comprehensive or nuanced picture. Low response rates, minimal follow-up capabilities, and a perceived lack of personalization can result in incomplete data that undermine effective decision making \cite{dillman2014internet,galesic2009effects,sahlqvist2011effect}. Many participants also see static questionnaires as impersonal and time-consuming, exacerbating survey fatigue and reducing the likelihood of thoughtful, candid feedback \cite{porter2005mail}.

Recent developments in Generative AI (GenAI) offer a promising alternative. Advances in large language models (LLMs) and conversational AI chatbots have paved the way for adaptive, interactive survey tools that address the shortcomings of traditional methods. A GenAI-powered chatbot can engage users in personalized, context-aware dialogues, promoting deeper reflection and richer responses than conventional surveys typically allow. Its capacity to ask real-time follow-up questions provides further clarity and nuance, leading to data that can better inform institutional initiatives.

In this paper, we introduce TigerGPT, an AI chatbot designed to assess campus climate. Unlike role-based or rigidly scripted survey tools, TigerGPT adapts to each user’s context—dynamically adjusting to ambiguous or erroneous inputs and offering flexible conversation flows. We hypothesize that this adaptive interface will yield deeper insights and higher user engagement than traditional methods. To test this hypothesis, we conducted a pilot study with students, collecting both quantitative metrics (e.g., satisfaction ratings) and qualitative feedback (e.g., open-ended comments, user reports). Our main contributions include:
\begin{itemize}
    \item {A New Adaptive Chatbot That Combines Role-Based Initialization with AI-Driven Dialogue.} TigerGPT  tailors initial prompts based on user roles and transcends rigid rule-based approaches by dynamically adapting to real-time inputs. This role-based method delivers both convenience and personalization, fostering deeper user engagement and more nuanced feedback.
    \item {Theory-Driven Engagement Approach.} Drawing on established principles of conversational design, TigerGPT employs empathetic cues, bolded questions, and user-driven topic selection to encourage sustained interaction. Sentiment analyses and user reports indicate that these strategies help create a welcoming environment and maintain user attention.    
    \item {Empirical Evidence of Positive User Satisfaction and Richer Feedback.} Feedback from students  shows that most participants reported at least moderate satisfaction with TigerGPT, often preferring it over traditional questionnaires. Their detailed, qualitative responses suggest the system successfully elicits more comprehensive user feedback.

\end{itemize}

The remainder of this paper is organized as follows: Section II reviews the literature on chatbot-based surveys and generative AI solutions. Section III presents TigerGPT’s design, methodology, and technical implementation.  Section IV reports our pilot study setup, evaluation results and analysis, emphasizing quantitative measures and qualitative insights drawn from participant essays. Section V concludes by summarizing key findings, discussing limitations, and future research directions.

\section{Related Work}
AI–powered chatbots are increasingly deployed in universities to assist with student learning and administrative tasks. By providing on-demand help with course registration, financial aid inquiries, and other processes, these systems can reduce faculty and staff workloads \cite{martinez2024ai, abbas2022online}. For instance, \cite{martinez2024ai} discusses a pilot at the European University of Madrid where a chatbot offering real-time information and emotional support led to more engaged, personalized learning experiences. Harvard’s CS50 used AI integration to nearly replicate a 1:1 teacher–student ratio, illustrating how carefully designed chatbot tools can support complex educational needs \cite{liu2024teaching}. In addition, studies highlight how chat-based interactions can improve communication among students, instructors, and campus resources, thereby lowering barriers to academic success \cite{abbas2022online, abbas2021university}.

The use of chatbots in survey research has gained traction as a way to boost engagement and collect richer data. Unlike static questionnaires, conversational surveys can adapt to user responses in real time, often prompting clarifications or in-depth follow-up questions \cite{xiao2020tell, tanwar2024opinebot}. This approach can alleviate survey fatigue and yield higher-quality feedback than traditional forms \cite{xiao2020tell, abbas2022online}. For example, \cite{xiao2020tell} found that participants offered more nuanced insights through chatbot-based surveys compared to a standard questionnaires. Similarly, \cite{tanwar2024opinebot} demonstrated how LLMs could tailor student feedback prompts, leading to greater user engagement and more comprehensive responses. 
However, there are concerns with this approach. A study by \cite{zarouali2024comparing} showed that some participants still preferred conventional web-based surveys, citing concerns over security or discomfort with automated systems. Likewise, \cite{njeguvs2021conversational} reported that while many users considered chatbot surveys “fun,” a considerable subset viewed human-led interactions as more trustworthy, underscoring sociocultural and trust-related challenges.

Early survey chatbots often relied on rule-based or flowchart-driven logic, which provided minimal adaptability \cite{thorat2020review, maeng2021designing}. Because these systems struggled with unanticipated inputs, any deviation from expected keywords or pathways could stall the conversation or result in irrelevant prompts \cite{chan2022challenges}. In contrast, recent AI-driven approaches use natural language processing (NLP) and LLMs to flexibly interpret user input and provide context-aware responses \cite{xiao2020tell, tanwar2024opinebot}. This adaptability proves especially beneficial in higher education, where participants may need personalized follow-ups or clarifications \cite{belhaj2021engaging, kim2019comparing}.

Nevertheless, AI chatbots present their own challenges, including user skepticism around data security and perceived impersonality \cite{zarouali2024comparing, njeguvs2021conversational}. While advanced models can reduce repetitive question fatigue and furnish real-time clarifications, they may struggle to balance open-ended adaptability with user trust—particularly when the survey topics involve sensitive issues like campus climate.

Overall, prior studies suggest that AI chatbots have the potential to significantly enhance user engagement and data quality in higher education survey contexts, compared to rigid rule-based approaches. Building on these insights, we developed TigerGPT, a GenAI chatbot that draws on established principles of conversational design and employs empathetic
cues, bolded questions, and user-driven topic selection.
It 
retains some role-based efficiency (e.g., collecting user role through quick clicks) but goes beyond static scripts by employing GenAI adaptiveness. This design allows users to access the most relevant survey templates with minimal typing effort, while still enabling the chatbot to handle ambiguous inputs through context-aware dialogue. The following section details how TigerGPT’s architecture integrates these role-based and open-ended strategies to provide a more personalized, real-time campus climate survey experience.

\begin{figure*}[t]
    \centering
    \includegraphics[width=0.95\linewidth]{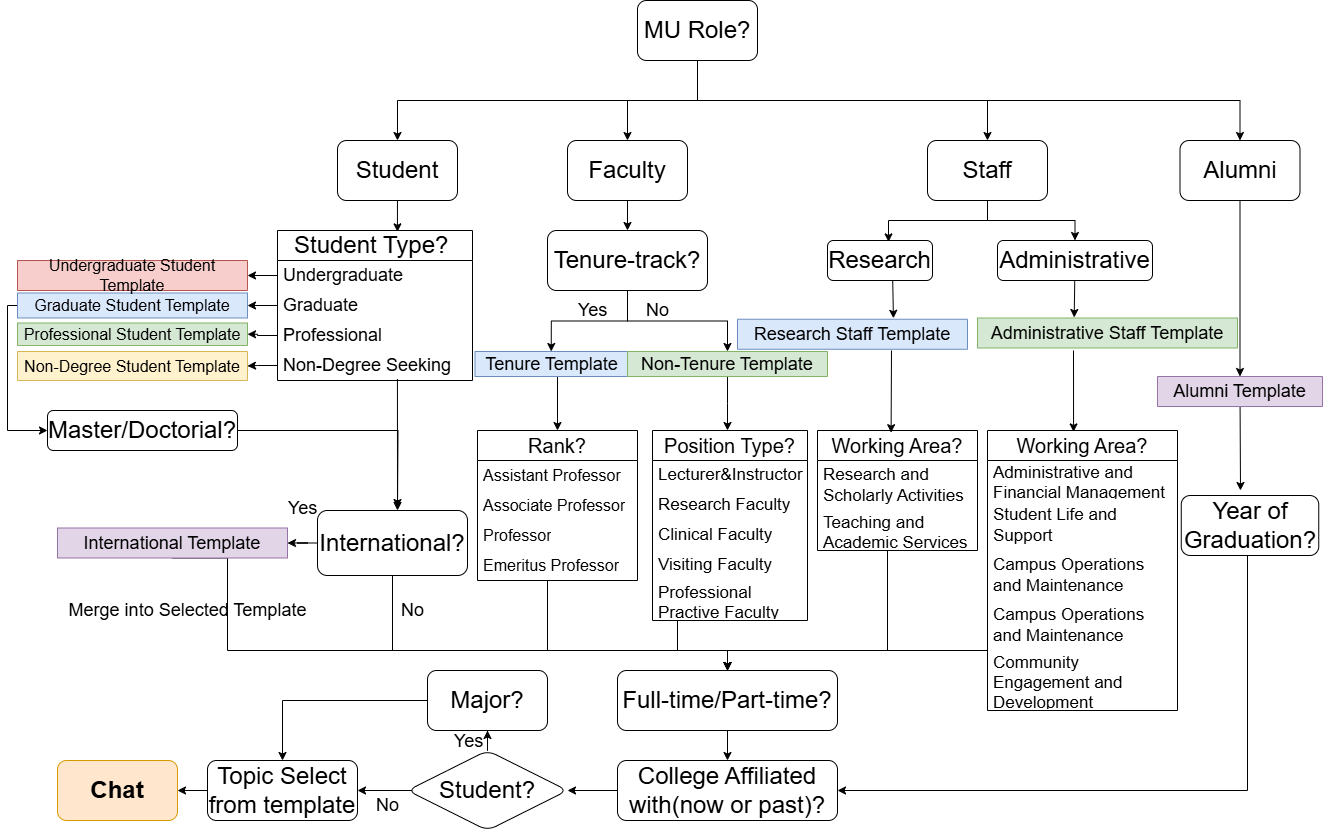} 
    \caption{{Overview of TigerGPT's initial role selection process.} This flowchart illustrates TigerGPT’s selection logic for campus climate surveys at the University of Missouri. Each user’s role (student, faculty, staff, or alumni) plus details such as a student’s type, a faculty member’s rank, or a staff member’s working area guides them to the most suitable survey template and topics.}
    \label{fig:flowchart}
\end{figure*}

\begin{figure}[t]
    \centering
        \includegraphics[width=1\linewidth]{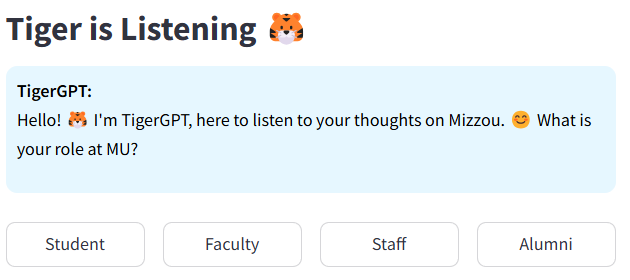}
    \caption{A screen shot of TigerGPT’s initial interface. A welcome screen prompting users to select their role at MU, as student, faculty, staff or alumni, to tailor subsequent interactions accordingly.}
    \label{fig:Preview}
\end{figure}


\section{TigerGPT: A New Chatbot for Adaptive Campus Climate Surveys}

\begin{figure}
    \centering
    \includegraphics[width=1\linewidth]{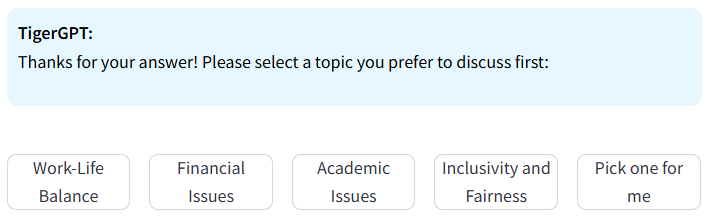}
    \caption{ TigerGPT offers users a variety of survey topics and an option of random selection.}
    \label{fig:subfig_multiple}
\end{figure}

TigerGPT is powered by advanced LLMs, such as OpenAI’s ChatGPT-4-turbo, and enables dynamic and context-aware conversations. As illustrated in Fig.~\ref{fig:flowchart}, the chatbot begins with a brief role selection process, selecting the roles of student, faculty, staff or alumni, along with details such as degree level, international status, and faculty track or working area. This concise setup helps users reduce typing and quickly load the most relevant survey template. Rather than confining the chatbot to a rigid script, however, the chatbot dynamically tailors subsequent questions and responses based on context and conversation history, capable of handling ambiguous or inconsistent inputs.  Fig.~\ref{fig:Preview} shows a screen shot of TigerGPT’s greetings and initial interface.
Fig.~\ref{fig:subfig_multiple} shows a screen shot of a variety of survey topics and an option of random selection.

TigerGPT uses {Streamlit} to implement its web interface  and {LangChain} to orchestrate calls to the OpenAI API, calling LLMs, such as ChatGPT-4-turbo, to generate responses in real time. Survey participants receive conversation flows that closely match their specific contexts while enjoying a smooth and engaging survey experience.

\begin{figure}[!t]
    \centering
        \includegraphics[width=\linewidth]{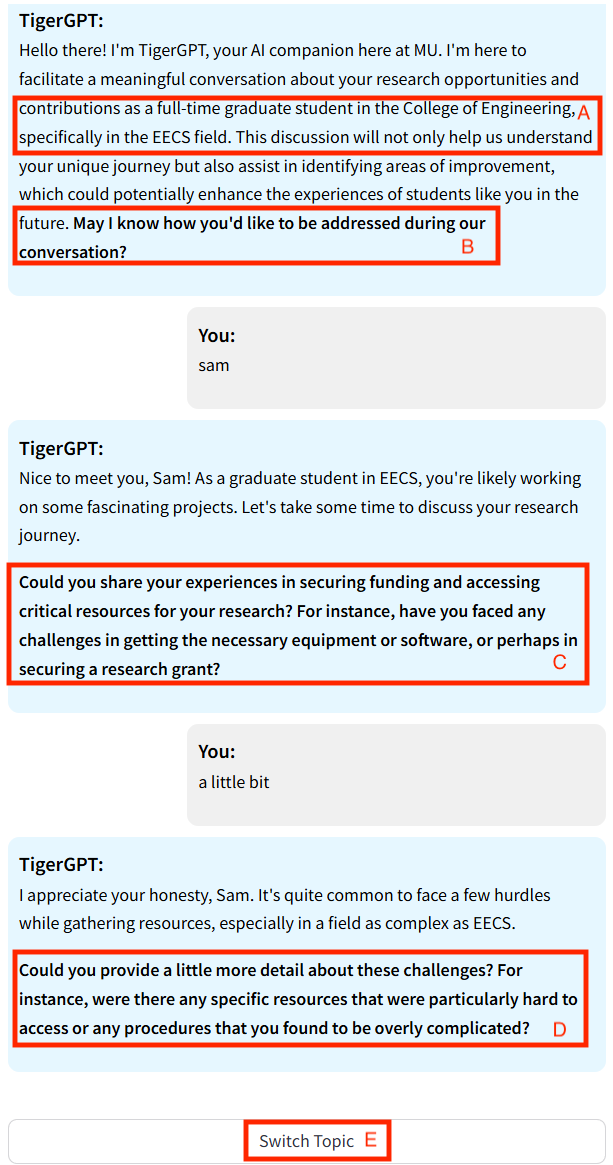}
    \caption{{Sample Conversation with TigerGPT.} This screenshot demonstrates five features: (A) the chatbot’s awareness of user information, (B) a prompt for the user’s preferred name, (C) a bold question supported by examples, (D) a follow-up request for more details, and (E) a “Switch Topic” button allowing the user to change the conversation’s direction.}
    \label{fig:mainFeatures-first}
\end{figure}

TigerGPT is designed based on established principles of
conversational design, employing personalization, empathetic
cues, bolded questions, and user-driven topic selection to
encourage sustained interaction and create a
welcoming environment and maintain user attention. Its key features are presented below.


\subsection{{Personalized Conversation}} Personalization in chatbots can significantly boost user engagement and satisfaction \cite{kocaballi2019personalization,chen2024recent}. In TigerGPT, each dialogue session is tailored to the user’s role, preferences, and contextual details (e.g., major, research focus), resulting in more targeted prompts and responses. An example of this personalization feature is shown in Fig.~\ref{fig:mainFeatures-first} (Label A).

Once the user makes their initial selections (e.g., role, degree program), TigerGPT stores these attributes in a short “user profile” prompt. This profile is then inserted into each conversation request so the chatbot remains aware of the user’s background. By merging the user profile with the active survey template, TigerGPT consistently delivers content that feels both relevant and personalized.

\subsection{{Asking for a Participant’s Preferred Name}} 
Using a participant’s chosen name is a personalization technique that fosters rapport and familiarity, making interactions feel more personal and inviting \cite{reghunath2021expression}. By greeting survey participants with their preferred name or referencing it throughout the conversation (see Fig.~\ref{fig:mainFeatures-first} (Label B)), TigerGPT creates a more human-like presence that can deepen engagement and overall satisfaction.

TigerGPT is prompted to introduce itself and ask for the participant’s preferred name at the  first interaction without asking any other questions. If a name is provided, the system stores it in session memory and integrates it into subsequent prompts (e.g., “Name, could you share more about…?”). If no name is given, TigerGPT defaults to a friendly, generic greeting (e.g., “Hey there!”), ensuring that all interactions remain welcoming and personalized, regardless of the user’s choice.

\subsection{{Clear, Bold Prompts with Examples}} Guided by principles of cognitive load reduction \cite{lenzner2010cognitive}, TigerGPT uses bolded questions and brief guiding examples to quickly draw the user’s attention to each query. \textcolor{red}{Explain what are bolded questions.} It is prompted to ask one open-ended question at a time, using double asterisks to make each query appear bold on-screen and underscore its importance. Each topic also includes a short guidance example illustrating the depth and style of detail users might provide. This approach clarifies expectations, encourages more thoughtful input, and ensures users fully grasp each query (see Fig.~\ref{fig:mainFeatures-first} (Label C)).

\subsection{{Follow-Up Prompts for Gathering Nuanced Information}} Studies on conversational interfaces show that well-placed follow-up questions encourage users to add nuance to their replies, often yielding richer, more detailed feedback \cite{pitts2007improving}. For instance, after a brief or unclear response, a chatbot might ask, “Could you tell me more about that?”, a technique that can lead to more in-depth answers than a single-question approach. TigerGPT adopts this strategy when user inputs appear too short or ambiguous, promoting deeper discussion and a fuller understanding of user needs (see Fig.~\ref{fig:mainFeatures-first} (Label~D)).

TigerGPT is specifically prompted to offer additional context or examples if a response seems insufficiently detailed. However, to avoid overwhelming users,  TigerGPT only requests further elaboration once per main topic, rather than after every reply. This balanced approach encourages users to expand on their answers without disrupting the conversational flow.

\subsection{{User Control via a “Switch Topic” Button}} TigerGPT provides a dedicated button that allows users to switch to a different survey topic they have not yet explored, aligning with usability principles that emphasize “user control and freedom.” By providing a “Switch Topic” button, TigerGPT lets participants gracefully redirect the dialogue and maintain a more user-centric interaction (see Fig.~\ref{fig:mainFeatures-first} (Label~E)). When this button is clicked, the system checks which topics have not been discussed and presents them to the user, ensuring participants can freely explore multiple survey areas without feeling constrained by a single line of questioning. This mechanism  improves the chatbot experience.

\begin{figure}[!t]
    \centering
        \includegraphics[width=\linewidth]{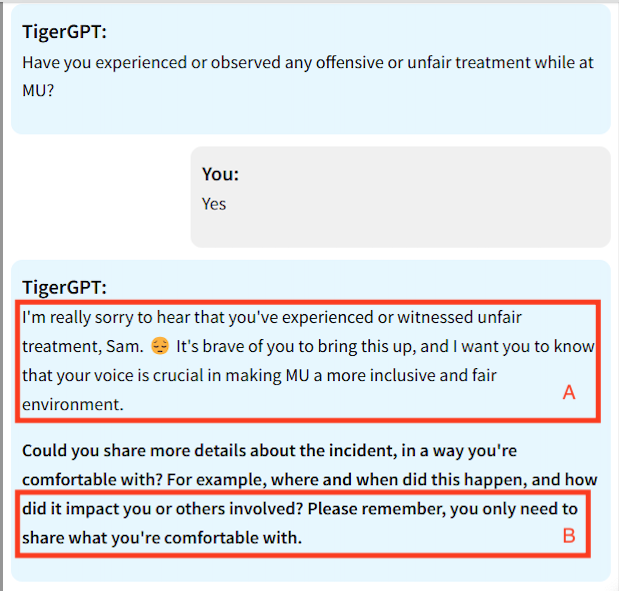}
    \caption{{Sample Conversation for a Sensitive Topic.} This example shows TigerGPT’s empathetic approach to user experiences of unfair treatment. In (A), the chatbot uses active listening and supportive language (with emojis) to validate the user’s feelings. In (B), it invites the user to share more details only at their comfort level, illustrating a respectful framework for discussing sensitive issues.}
    \label{fig:mainFeatures-second}
\end{figure}

\subsection{{Active Listening, Empathetic Responses, and Emoji Implementation}} Active listening and empathetic language foster trust and engagement by showing genuine concern for users’ experiences. Research suggests that empathic chatbot responses significantly enhance user satisfaction \cite{zhou2025building}, while well-chosen emojis convey additional warmth and emotional cues, contributing to a more human-like conversation \cite{bai2019systematic}. Together, these elements encourage users to feel heard and comfortable discussing personal topics (see Fig.~\ref{fig:mainFeatures-second} (Label A)).

To achieve this, TigerGPT maintains a supportive, non-judgmental tone throughout each interaction, acknowledging users’ courage in sharing sensitive information. When users express distress or other strong emotions, the system responds with empathetic phrases. At strategic points, emojis are used to visually reinforce these emotional nuances, further enhancing user engagement and rapport.

\subsection{{Encouraging Comfortable Information Sharing}}
Openly discussing sensitive or personal matters can be challenging for participants, especially if they perceive judgment or pressure \cite{melville2016conducting}. The concept of self-disclosure in social psychology emphasizes that individuals are more likely to share personal information when they trust their conversational partner and retain control over how much to reveal \cite{wheeless1977measurement}. Additionally, respect for autonomy in user-centered design \cite{kim2008keeping} underscores the importance of allowing participants to set their own boundaries, thereby creating a safer environment for disclosure (see Fig.~\ref{fig:mainFeatures-second} (Label B)).

TigerGPT  maintains a supportive, non-judgmental tone throughout each interaction—especially when topics become sensitive. If the user shows hesitation or distress, the chatbot adjusts its questions accordingly, refraining from further probing unless the user explicitly indicates a willingness to continue. Where relevant, TigerGPT also provides information on support resources, avoiding any demands for personal data. This balanced approach encourages deeper sharing while preserving participants’ sense of security.

\subsection{{Providing Multiple Topic Options for Personalized Guidance}} To enhance user autonomy and relevance, TigerGPT presents a set of topics aligned with each individual’s profile, along with an option to pick one randomly. This design choice allows participants to guide the conversation toward areas they find most pertinent or interesting, while also offering a more free-form alternative for those who prefer spontaneity (see Fig.~\ref{fig:subfig_multiple}). By empowering users to choose the next topic, TigerGPT fosters greater engagement and ensures each session remains tailored to personal preferences.

\section{Evaluation}
To evaluate the performance of TigerGPT, we recruited undergraduate students in a psychology class from the University of Missouri to test it. We collected data from two primary sources: (1) a traditional questionnaire-based feedback survey conducted immediately following each TigerGPT survey, and (2) written evaluation reports submitted by student evaluators.

\begin{figure*}[!t]
    \centering
        \includegraphics[width=1.0\linewidth]{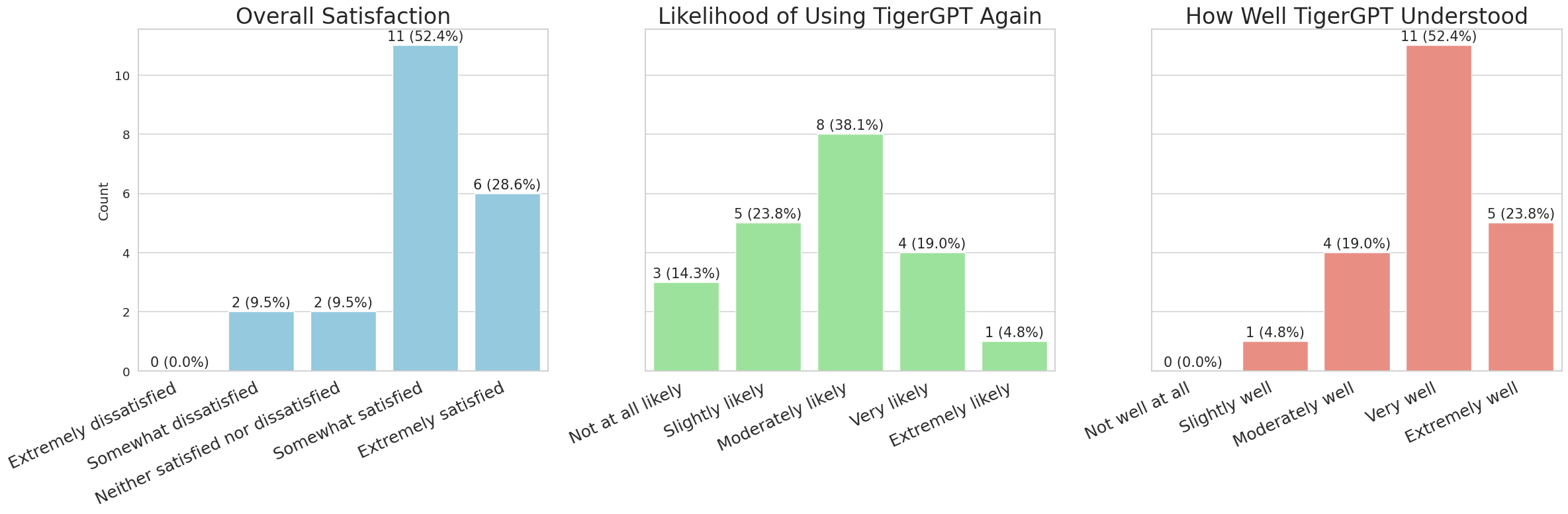}
    \caption{{User Opinions towards TigerGPT.} These three bar charts show participants’ overall satisfaction, likelihood of using TigerGPT again, and how well the system understood their queries.}
    \label{fig:opinion}
    \footnotesize
\end{figure*}

\begin{figure}[!t]
    \centering
        \includegraphics[width=0.7\linewidth]{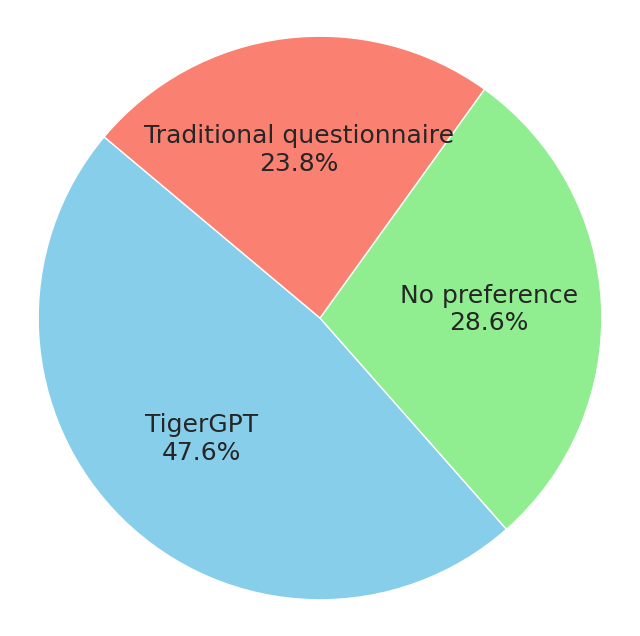}
    \caption{{Participant Preference Chart.} This pie chart shows how participants prefer TigerGPT, a traditional questionnaire, or neither.}
    \label{fig:preference}
\end{figure}

\subsection{{User Feedback Survey Results}}
A total of 21 participants completed the feedback survey, summarized in Fig.~\ref{fig:opinion}. Approximately 81\% reported being at least “somewhat satisfied,” with 28.6\% indicating they were “extremely satisfied,” and only 9.5\% expressing dissatisfaction. Nearly 62\% noted they were “moderately likely” or more inclined to use the system again, while a smaller proportion showed lower willingness. Regarding perceived comprehension, around 76\% felt their input was understood “very well” or “extremely well,” suggesting that most respondents found the tool’s responses suitably tailored.

As shown in Fig.~\ref{fig:preference}, about half of the participants preferred TigerGPT, roughly a quarter preferred traditional questionnaires, and the remainder had no strong preference. Overall, these findings indicate a generally positive reception, tempered by some uncertainty about whether this approach offers decisive advantages over traditional survey methods.

\begin{figure*}[!t]
    \centering
        \includegraphics[width=0.9\linewidth]{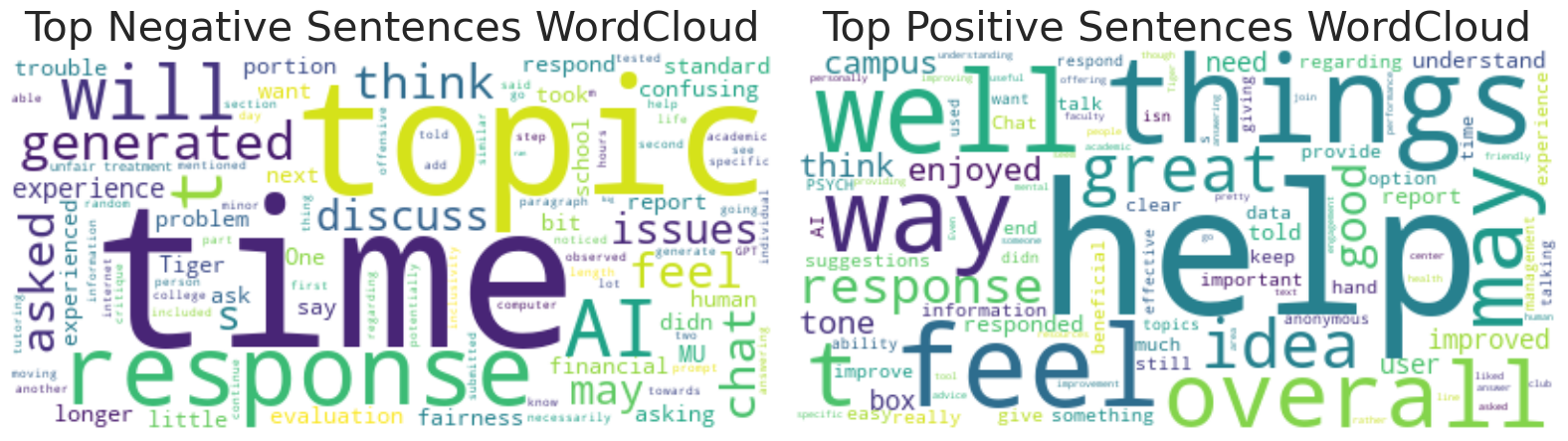}
    \caption{{Word Cloud Visualization of User Feedback.} The left cloud highlights frequently mentioned words from negative feedback (e.g., time, topic, response), while the right cloud shows terms from positive feedback (e.g., help, overall, things). These visuals provide a quick overview of users’ main concerns and appreciation points.}
    \label{fig:wordcloud}
    \footnotesize
\end{figure*}

\begin{figure*}[!t]
    \centering
        \includegraphics[width=1.0\linewidth]{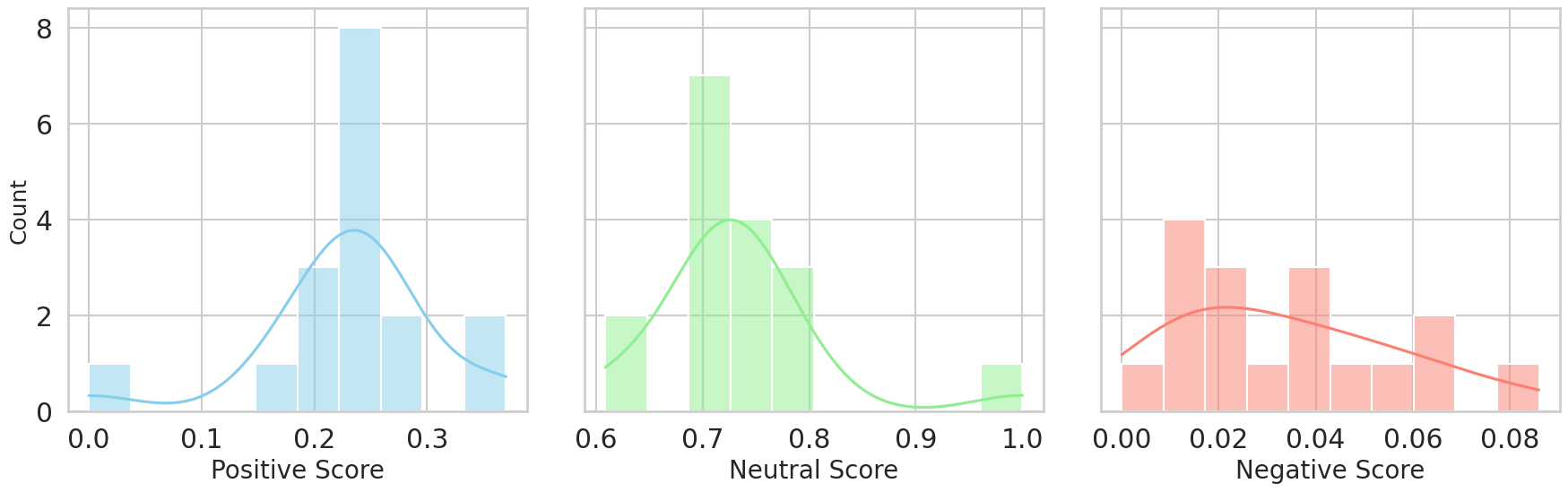}
    \caption{{Distribution of Positive, Neutral, and Negative Sentiment Scores.} This figure displays three histograms illustrating how frequently users’ feedback falls into varying degrees of positive, neutral, and negative sentiment. Each histogram shows the number of responses (y-axis) at different sentiment score ranges (x-axis), with a smoothing curve overlaid to highlight the general distribution trend.}
    \label{fig:sentiment distribution}
    \footnotesize
\end{figure*}

\subsubsection{\textbf{Evaluation Report Results}}

A total of 17 students submitted written reports discussing both promising aspects and areas of concern. Figure~\ref{fig:wordcloud} shows word clouds derived from their most positive and negative statements. In the positive word cloud, terms such as \emph{help}, \emph{overall}, and \emph{things} appear frequently, reflecting an appreciation for TigerGPT’s assistance and a generally favorable user experience. Meanwhile, in the negative word cloud, words like \emph{time}, \emph{topic}, and \emph{response} stand out, suggesting frustration with occasional slow responses or limited topic coverage. This contrast illustrates the dual nature of user sentiment: while many enjoyed TigerGPT’s interactivity and friendly tone, some felt it could better handle broader or more flexible discussion prompts.

Figure~\ref{fig:sentiment distribution} further illustrates how users’ written feedback distributes across \emph{positive}, \emph{neutral}, and \emph{negative} sentiment scores, based on VADER (Valence Aware Dictionary and sEntiment Reasoner). Each histogram shows the number of responses (y-axis) in different score intervals (x-axis), with a smoothing curve to highlight overall trends. Notably, positive scores cluster around 0.2–0.3, neutral scores center near 0.7, and negative scores remain low (generally below 0.08). VADER calculates these values by analyzing lexical cues and punctuation, then produces an overall polarity (the \emph{compound} score) alongside distinct positive, neutral, and negative metrics.

Table~\ref{tab:sentiment_stats} provides descriptive statistics: the mean compound score stands at 0.93, confirming the predominantly positive tone seen in the word clouds. Although some comments addressed issues like slow response times or narrow topic options—reflected in minor negative scores—the overall distribution indicates that students generally found TigerGPT both helpful and engaging. Taken together, these results suggest that while most users had a favorable experience, improvements to performance and topic breadth could further enhance satisfaction.

\input{table1}

Building on these sentiment findings, many students appreciated the system’s {flexible, user-friendly design}. They highlighted features like its {cross-platform interface} supporting smartphones, tablets, and computers, and {“pick one for me” prompts}, which helped them explore new topics without confusion. The inclusion of {bolded key questions} was particularly beneficial for those with attention difficulties; one student mentioned it helped them stay focused during the conversation. Students also felt {safe and anonymous} when giving feedback, promoting more direct and detailed responses and making the experience {more engaging than a typical survey}.

Regarding content, participants praised the {wide range of relevant topics}, encompassing academic life, financial aid, work–life balance, and campus inclusivity. They also enjoyed the chatbot’s {human-like elements}; for instance, it addressed them by name and major, used an {enthusiastic tone} (often with emojis), and offered {friendly affirmations} that created a warm, conversational atmosphere. Subtle {university branding} (e.g., university icons) reinforced the campus feel without overwhelming the user. Lastly, several students saw {potential for deeper academic integration}, suggesting the chatbot could link to the university’s scheduling system or relevant resource pages, reducing the need for in-person visits and offering a convenient way to access campus support.

While participants generally appreciated the system, they also identified areas for potential improvement. For example, technical stability enhancements could address occasional slow or inconsistent response times, as well as the need to re-enter responses after a page reload. Diversifying language choices might reduce repetitive phrases (e.g., “I see” or “That’s amazing”) and enable more in-depth follow-ups. Some users also proposed broadening the range of topics, covering a larger and more diverse population, and allowing a more flexible conversation flow so predefined prompts would not limit discussions of pressing issues. Finally, several students wished for more detailed feedback from the system, suggesting that deeper insights would better capture their opinions and enrich the overall experience.

\section{Conclusion}
In this paper, we introduced TigerGPT, an AI chatbot designed to make campus climate surveys more personalized and engaging. Overall, participants testing the chatbot responded positively to TigerGPT’s conversational flow, empathetic responses, visual aids, and flexible topic selection. About half preferred TigerGPT over a traditional questionnaire, noting how features like addressing users by name, empathetic cues, and offering friendly affirmations made the experience feel warmer and more human.

Despite these encouraging results, participants identified several areas for improvement. Some reported slow or inconsistent response times and repetitive language (e.g., frequent “I see” statements), while others suggested broadening the range of conversation flows. Additionally, while TigerGPT currently uses role-selection to quickly personalize survey prompts, its scope of topics remains relatively narrow, potentially overlooking certain campus issues.

In our future work, we aim to develop a robust, intuitive, and adaptable survey platform that continues to build trust, encourage candid feedback, and ultimately guide meaningful campus improvements. Future research will investigate how chatbot-led surveys affect response quality and user satisfaction across more diverse academic settings. Advanced user modeling may reduce reliance on explicit role selection and enable a truly free-flowing, AI-driven conversation.

\section*{Acknowledgment}
ChatGPT was used to revise the writing to improve the spelling, grammar, and overall readability.

\bibliographystyle{IEEEtran}
\bibliography{bibi_chat}

\end{document}

%% file: table1.tex
\begin{table}[!t]
\caption{\textbf{VADER Sentiment Score Statistics.} Descriptive statistics (mean, standard deviation, minimum, and maximum) for compound, positive, neutral, and negative sentiment scores across 17 user reports. The average compound score of 0.93 suggests an overall positive sentiment trend.}
\label{tab:sentiment_stats}
\centering
\begin{tabular}{lcccc}
\hline
 & \textbf{Compound} & \textbf{Positive} & \textbf{Neutral} & \textbf{Negative} \\
\hline
\textbf{Mean} & 0.93 & 0.23 & 0.74 & 0.03 \\
\textbf{SD}   & 0.24 & 0.08 & 0.08 & 0.02 \\
\textbf{Min}  & 0.00 & 0.00 & 0.61 & 0.00 \\
\textbf{Max}  & 1.00 & 0.37 & 1.00 & 0.09 \\
\hline
\multicolumn{5}{l}{\textbf{Number of reports:} 17}\\
\multicolumn{5}{l}{\textbf{Average Compound Score:} 0.93}\\
\hline
\end{tabular}
\end{table}